\documentclass[conference]{IEEEtran}
\IEEEoverridecommandlockouts
\usepackage{cite}
\usepackage{amsmath,amssymb,amsfonts}
\usepackage{algorithmic}
\usepackage{graphicx}
\usepackage{textcomp}
\usepackage{xcolor}
\usepackage{algorithm}      % for writing algorithms
\usepackage[utf8]{inputenc}
\usepackage{pgfplots}
\usepackage[acronym, shortcuts]{glossaries}
\usepackage{stylesheet}
\newacronym{wasn}{WASN}{wireless acoustic sensor network}
\newacronym{danse}{DANSE}{distributed adaptive node-specific signal estimation}
\newacronym{tidanse}{TI-DANSE}{topology-independent distributed adaptive node-specific signal estimation}
\newacronym{tigevddanse}{TI-GEVD-DANSE}{topology-independent GEVD-based distributed adaptive node-specific signal estimation}
\newacronym{ti_or_gevddanse}{TI-(GEVD-)DANSE}{topology-independent (GEVD-based) distributed adaptive node-specific signal estimation}
\newacronym{gevddanse}{GEVD-DANSE}{GEVD-based distributed adaptive node-specific signal estimation}
\newacronym{gevd}{GEVD}{generalized eigenvalue decomposition}
\newacronym{lmmse}{LMMSE}{linear minimum mean square error}
\newacronym{mwf}{MWF}{multichannel Wiener filter}
\newacronym{gevdmwf}{GEVD-MWF}{GEVD-based multichannel Wiener filter}
\newacronym{scm}{SCM}{spatial covariance matrix}
\newacronym{snr}{SNR}{signal-to-noise ratio}
\newacronym{mse}{MSE}{mean square error}

\DeclareUnicodeCharacter{2212}{−}
\usepgfplotslibrary{groupplots,dateplot}
\usetikzlibrary{patterns,shapes.arrows,external}
\pgfplotsset{compat=newest}

\def\BibTeX{{\rm B\kern-.05em{\sc i\kern-.025em b}\kern-.08em
    T\kern-.1667em\lower.7ex\hbox{E}\kern-.125emX}}
\begin{document}

\title{Topology-independent GEVD-based\\distributed adaptive node-specific signal estimation \\in ad-hoc wireless acoustic sensor networks\\

\thanks{
    This research was carried out at the ESAT Laboratory of KU Leuven, in the frame of Research Council KU Leuven C14-21-0075 ``A holistic approach to the design of integrated and distributed digital signal processing algorithms for audio and speech communication devices'', and was supported by the European Union's Horizon 2020 research and innovation programme under the Marie Skłodowska-Curie grant agreement No. 956369: `Service-Oriented Ubiquitous Network-Driven Sound — SOUNDS'. The scientific responsibility is assumed by its authors. This paper reflects only the authors' views and the Union is not liable for any use that may be made of the contained information. 
}
}

\author{\IEEEauthorblockN{Paul Didier, Toon van Waterschoot, Marc Moonen}
\IEEEauthorblockA{\textit{KU Leuven, Department of Electrical Engineering (ESAT),}\\
\textit{STADIUS Center for Dynamical Systems, Signal Processing and Data Analytics, Belgium} \\
Email: {paul.didier, toon.vanwaterschoot, marc.moonen}@esat.kuleuven.be}}

\maketitle

\begin{abstract}
    A low-rank approximation-based version of the 
    topology-independent distributed adaptive node-specific signal estimation (TI-DANSE) algorithm is introduced, using a generalized eigenvalue decomposition (GEVD) for application in ad-hoc wireless acoustic sensor networks. This TI-GEVD-DANSE algorithm as well as the original TI-DANSE algorithm exhibit a non-strict convergence, which can lead to numerical instability over time, particularly in scenarios where the estimation of accurate spatial covariance matrices is challenging. An adaptive filter coefficient normalization strategy is proposed to mitigate this issue and enable the stable performance of \mbox{TI-(GEVD-)DANSE}. The method is validated in numerical simulations including dynamic acoustic scenarios, demonstrating the importance of the additional normalization.
\end{abstract}


\section{Introduction}
In recent years, the ever-increasing ubiquity of multi-microphone devices capable of exchanging and processing acoustic signals has motivated the development of distributed audio signal processing algorithms.
As opposed to traditional localized microphone arrays, distributed systems do not rely on a fusion center; they instead leverage the computing capacities of each device (i.e., each node) in \glspl{wasn}. Distributed systems are typically able to use signals spanning a large acoustic area while maintaining a high degree of flexibility in their physical design~\cite{bertrandApplications, boukercheDesign}.

The tasks of distributed algorithms may be categorized depending on the estimated quantity~\cite{wuReview}.
Here, the focus is set on signal estimation for applications that require the retrieval of entire (possibly multichannel) signals of interest which may be node-specific and non-stationary, e.g., for noise reduction in speech enhancement tasks.
In a fully connected \gls{wasn}, one may let all nodes transmit all their local sensor signals to all other nodes, effectively corresponding to a centralized case. For obvious reasons, this strategy suffers from an inefficient usage of communication bandwidth. Instead, it has been shown that the nodes can exchange fused versions of their local sensor signals while retaining the same performance as if they were transmitting all of their sensor signals. This is a core idea in the \gls{danse} algorithm~\cite{bertrandDistributed}, which is considered in this paper.

The \gls*{danse} algorithm is iterative and converges towards the centralized \gls{lmmse} optimum. It operates in fully connected \glspl{wasn}, where every node can communicate with every other node. However, in many practical applications, there is no guarantee that the network topology will be fully connected or be static over time (e.g., due to link failures).
A solution to this is provided by the topology-independent (TI) \gls{danse} algorithm (\glsunset{tidanse}\mbox{\gls{tidanse}})~\cite{szurleyTopology}, which allows a new tree to be pruned from the ad-hoc topology at any algorithm iteration while retaining convergence, thus being robust to dynamic topologies.

The \gls*{danse} and \gls*{tidanse} algorithms rely at their core on the computation of a \gls{mwf}.
A significant performance improvement can be obtained when the number of latent target sources, or an estimate thereof, is known a priori -- a reasonable assumption in many applications. The rank of the desired signal \gls{scm} can be set equal to that number via a \gls{gevd}. It has been shown that such \gls{gevd}-based low-rank approximation of the \gls{mwf} (\glsunset{gevdmwf}\mbox{\gls{gevdmwf}}) indeed outperforms the unconstrained \gls{mwf}, particularly in challenging signal-to-noise ratio conditions~\cite{serizelLow}. Although the concept has been applied to \gls{danse} in~\cite{hassaniGEVD}, leading to the \glsunset{gevddanse}\gls{gevddanse} algorithm, its potential for \mbox{\gls{tidanse}} has remained unexplored.

The contribution of this paper is two-fold. First, we introduce the \glsunset{tigevddanse}\mbox{\gls{tigevddanse}} algorithm, which can operate in any topology and incorporates the advantages of a \mbox{\gls{gevdmwf}}. Second, we address the observed non-strict convergence of the \mbox{\gls{tigevddanse}} algorithm, also observed in the original \mbox{\gls{tidanse}} algorithm, which allows the \gls{lmmse} optimum to be reached even though the filters themselves diverge.
To this end, we propose a filter coefficient normalization strategy which stabilizes the behavior of \glsunset{ti_or_gevddanse}\mbox{\gls{ti_or_gevddanse}} even when signal statistics are estimated on the fly.

This paper is organized as follows. The problem statement is given in~Section~\ref{sec:problemStatement}, defining the signal model and the centralized \gls{mwf} solution. The \mbox{\gls{tidanse}} algorithm is reviewed in~Section~\ref{sec:TIDANSE}. The \mbox{\gls{tigevddanse}} algorithm is presented in~Section~\ref{sec:TIGEVDDANSE}. The proposed normalization strategy addressing the stability of \mbox{\gls{ti_or_gevddanse}} is presented in~Section~\ref{sec:improvedRobustness}. Numerical experiments are provided in~Section~\ref{sec:numericalExperiments}. Finally, conclusions are given in~Section~\ref{sec:conclusion}.

%%%%%
\section{Problem Statement}\label{sec:problemStatement}

We consider a \gls{wasn} consisting of $K$ nodes, where node $k$ has $M_k$ sensors ($k\in\mathcal{K}:=\{1,...,K\}$) such that the total number of sensors is $M=\sum_{k\in\mathcal{K}}M_k$. The signals are assumed to be complex-valued to allow representation of, e.g., processing in a particular bin of a filter bank. The acoustic scenario includes $S$ desired sources whose latent signals at time $t$ are grouped in $\hat{\mathbf{s}}[t]\in\mathbb{C}^{S\times1}$, and $S_n$ noise sources with latent signals $\hat{\mathbf{n}}[t]\in\mathbb{C}^{S_n\times1}$. The sensor signals at all nodes are stacked in $\mathbf{y}[t]\in\mathbb{C}^{M\times 1}$, which is modeled as:

\begin{equation}\label{eq:signalmodel}
    \mathbf{y}[t] = \mathbf{s}[t] + \mathbf{n}[t]
    = \mathbf{A}\hat{\mathbf{s}}[t] + \mathbf{B}\hat{\mathbf{n}}[t] + \mathbf{q}[t],
\end{equation}

\noindent
where $\mathbf{n}[t]$ is the noise component, $\mathbf{A}\in\mathbb{C}^{M\times S}$ and $\mathbf{B}\in\mathbb{C}^{M\times S_n}$ are the steering matrices of the desired signal and the noise, respectively, and $\mathbf{q}[t]$ is uncorrelated (thermal) noise. Time indices $[t]$ are omitted from here on for conciseness.
Each vector in~(\ref{eq:signalmodel}) can be partitioned as 
$\mathbf{y} = [\mathbf{y}_1^\T\dots \mathbf{y}_K^\T]^\T$, where
$\cdot^\T$ denotes the transpose operator and $\mathbf{y}_k\in\mathbb{C}^{M_k\times 1}$ are the sensor signals of node $k$. Each node strives to estimate its own $J$-channel desired signal $\mathbf{d}_k = \mathbf{E}_{kk}^\T\mathbf{s}_k$ where $\mathbf{E}_{kk}$ is an $M_k\times J$ selection matrix.
A node-specific \gls{lmmse} estimation problem is considered, where the optimal filter matrix $\mathbf{W}_k\in\mathbb{C}^{M\times J}$ used to estimate $\mathbf{d}_k$ from $\mathbf{y}$ is defined as:

\begin{equation}\label{eq:mmseCentr}
    \mathbf{W}_k = \underset{\mathbf{W}}{\mathrm{arg\,min}}\:
    \mathbb{E}\left\{
        \left\|
            \mathbf{d}_k - \mathbf{W}^\Her\mathbf{y}
        \right\|^2_2
    \right\},
\end{equation}

\noindent
where $\mathbb{E}\{\cdot\}$, $\|\cdot\|_2$, and $\cdot^\Her$ denote the expected value operator, the Euclidean norm, and the Hermitian operator, respectively.
The solution of (\ref{eq:mmseCentr}) is the well-known \gls{mwf}:

\begin{equation}
    \mathbf{W}_k=\left(\mathbf{R}_{\mathbf{yy}}\right)^{-1}\mathbf{R}_{\mathbf{ss}}\mathbf{E}_{k},
\end{equation}

\noindent
with the \glspl{scm} $\mathbf{R}_{\mathbf{yy}}=\mathbb{E}\{\mathbf{yy}^\Her\}$ and $\mathbf{R}_{\mathbf{ss}}=\mathbb{E}\{\mathbf{ss}^\Her\}$ and $\mathbf{E}_{k}$ the $M\times J$ selection matrix extracting $\mathbf{d}_k$ from $\mathbf{s}$. In practice, under the assumption that noise and desired signal components are uncorrelated, $\mathbf{R}_{\mathbf{ss}}$ may be estimated as $\mathbf{R}_{\mathbf{yy}} - \mathbf{R}_{\mathbf{nn}}$, where $\mathbf{R}_{\mathbf{nn}}=\mathbb{E}\{\mathbf{nn}^\Her\}$.
In real-world applications, \glspl{scm} must be estimated in an online fashion.
This can be achieved via, e.g., exponential averaging with a forgetting factor $0<\beta<1$ as:

\begin{equation}\label{eq:scmsOnline}
    \begin{split}
        \mathbf{R}_{\mathbf{y}\mathbf{y}}[t]
        &=
        \beta
        \mathbf{R}_{\mathbf{y}\mathbf{y}}[t-1]
        + (1 - \beta)
        \mathbf{y}[t]\mathbf{y}^\Her[t],\\
        \mathbf{R}_{\mathbf{n}\mathbf{n}}[t]
        &=
        \beta
        \mathbf{R}_{\mathbf{n}\mathbf{n}}[t-1]
        + (1 - \beta)
        \mathbf{n}[t]\mathbf{n}^\Her[t],
    \end{split}
\end{equation}

\noindent
where the same symbols are used for estimated and true quantities for simplicity of notation.
In practice, $\mathbf{n}[t]$ may not be directly available and can be extracted via, e.g., an activity detector exploiting the ON-OFF structure of speech-like target signals~\cite{bertrandDistributed,zhaoModel}.
The desired signal estimate is finally obtained as $\hat{\mathbf{d}}_k = \mathbf{W}_k^\Her\mathbf{y}$.

In practice, making $\mathbf{y}$ available requires either a fusion center or the exchange of $M_k$-dimensional signals between the nodes of a fully connected \gls{wasn}.

%%%%%%
\section{\gls*{tidanse}}\label{sec:TIDANSE}

The \gls{danse} algorithm~\cite{bertrandDistributed} may be used to reach the centralized solution of~(\ref{eq:mmseCentr}) while substantially reducing the communication bandwidth requirements. However, it requires a static and fully connected \gls{wasn} topology, which is rarely the case in practice.
In an ad-hoc, possibly time-varying \gls{wasn} topology, one may instead employ the \mbox{\gls{tidanse}} algorithm~\cite{szurleyTopology} which is reviewed in this section.

The \mbox{\gls{tidanse}} algorithm is iterative with iteration index $i$, i.e., it asymptotically converges towards the centralized \gls*{lmmse} solution. At any iteration, the \gls{wasn} topology can be pruned to a new tree using, e.g., Prim's algorithm. One node is chosen as the updating node, a natural choice being the root of the tree. The index of the updating node is set to cycle through all $k\in\mathcal{K}$ in a round-robin fashion, such that all nodes have updated once after $K$ iterations.
Each node $k$ defines its so-called fusion matrix $\mathbf{P}_k^i\in\mathbb{C}^{M_k\times J}$ and uses it to compute a $J$-channel fused version of its local sensor signals $\mathbf{z}_k^i = \mathbf{P}_k^{i\Her}\mathbf{y}_k$. All fused signals in the \gls{wasn} are summed up via a sequence of partial in-network signal summations from the leaf nodes towards the root node. Once the full $J$-dimensional in-network sum $\boldsymbol{\eta}^i = \sum_{k\in\mathcal{K}}\mathbf{z}_k^i$ is built at the root node, it is flooded back towards the leaf nodes. Through this sequence of operations, each node has access to $D_k = M_k+J$ signals grouped in an observation vector:

\begin{equation}\label{eq:yTildeTI}
    \tilde{\mathbf{y}}_k^i = [\mathbf{y}_k^\T\:|\: \boldsymbol{\eta}_{-k}^{i\T}]^\T
    \:\:\:\text{where}\:\:\:
    \boldsymbol{\eta}_{-k}^{i} = \boldsymbol{\eta}^{i} - \mathbf{z}_k^i = \sum_{q\in\mathcal{K}\backslash\{k\}}\mathbf{z}_q^i.
\end{equation}

The objective of node $k$ is then to solve its own node-specific \gls*{lmmse} problem to estimate $\mathbf{d}_k$ from $\tilde{\mathbf{y}}_k^i$:

\begin{equation}\label{eq:mmseTIDANSE}
    \tilde{\mathbf{W}}_k^{i+1} = \underset{\mathbf{W}}{\mathrm{arg\,min}}\:
    \mathbb{E}\left\{
        \left\|
            \mathbf{d}_k - \mathbf{W}^\Her\tilde{\mathbf{y}}_k^i
        \right\|^2_2
    \right\},
\end{equation}

\noindent
which, as in the centralized case, is solved by an \gls{mwf} as:

\begin{equation}\label{eq:mwf_danse}
    \tilde{\mathbf{W}}_k^{i+1} = (
        \tilde{\mathbf{R}}_{\mathbf{y}_k\mathbf{y}_k}^i
    )^{-1}
    \tilde{\mathbf{R}}_{\mathbf{s}_k\mathbf{s}_k}^i\tilde{\mathbf{E}}_{k},
\end{equation}

\noindent
with the \glspl{scm} $\tilde{\mathbf{R}}_{\mathbf{y}_k\mathbf{y}_k}^i = \mathbb{E}\{\tilde{\mathbf{y}}_k^i\tilde{\mathbf{y}}_k^{i\Her}\}$ and $\tilde{\mathbf{R}}_{\mathbf{s}_k\mathbf{s}_k}^i = \mathbb{E}\{\tilde{\mathbf{s}}_k^i\tilde{\mathbf{s}}_k^{i\Her}\}$, $\tilde{\mathbf{y}}_k^i = \tilde{\mathbf{s}}_k^i + \tilde{\mathbf{n}}_k^i$, and $\tilde{\mathbf{E}}_{k} = [\mathbf{E}_{kk}^\T \:|\: \mathbf{0}_{J\times J}]^\T$.
A partitioning of $\tilde{\mathbf{W}}_k^i$ is defined here as $[\mathbf{W}_{kk}^{i\T} \:|\: \mathbf{G}_{k}^{i\T}]^\T$, where $\mathbf{W}_{kk}^{i}\in\mathbb{C}^{M_k\times J}$ is applied to the $\mathbf{y}_k$ while $\mathbf{G}_{k}^{i}\in\mathbb{C}^{J\times J}$ is applied to $\boldsymbol{\eta}_{-k}^i$.

In~\cite{szurleyTopology}, the convergence and optimality of \mbox{\gls{tidanse}} is proved when the fusion rule is $\mathbf{P}_k^i = \mathbf{W}_{kk}^i(\mathbf{G}_{k}^i)^{-1}$, which completes the algorithm definition. It can be noted that the $(\mathbf{G}_k^i)^{-1}$ term serves to decouple the contributions of each individual node in $\boldsymbol{\eta}^i$. The \mbox{\gls{tidanse}} definition of $\mathbf{P}_k^i$ is a fundamental difference with that used in the \gls{danse} algorithm, where the $(\mathbf{G}_k^i)^{-1}$ term is omitted~\cite{bertrandDistributed}.
As $i\to\infty$, the \mbox{\gls{tidanse}} algorithm converges to the centralized solution to~(\ref{eq:mmseCentr}) while reducing the amount of information exchanged between nodes with respect to the centralized case, as well as to \gls{danse}. The algorithm assumes that the nodes update sequentially, in a round-robin fashion. Simultaneous or asynchronous node-updating strategies are not considered here.

%%%%%
\section{\mbox{\gls{tigevddanse}}}\label{sec:TIGEVDDANSE}

The \gls{mwf} solving~(\ref{eq:mmseCentr}) depends on $\mathbf{R}_\mathbf{ss}$, which is a rank-$S$ matrix since $\mathbf{R}_\mathbf{ss} = \mathbf{A}\mathbb{E}\{\hat{\mathbf{s}}\hat{\mathbf{s}}^\Her\}\mathbf{A}^\Her$ (cf.~(\ref{eq:signalmodel})). However, in practice, $\mathbf{R}_\mathbf{ss}$ is not directly available and must be estimated, for instance as $\mathbf{R}_\mathbf{ss} = \mathbf{R}_\mathbf{yy} - \mathbf{R}_\mathbf{nn}$, often resulting in %an estimated matrix with
a rank greater than $S$. In many practical applications such as speech enhancement, $S$ is known a priori or can be well-estimated. Constraining the rank of $\mathbf{R}_\mathbf{ss}$ can then be done via a \gls{gevd} of the pencil $\{\mathbf{R}_\mathbf{yy}, \mathbf{R}_\mathbf{nn}\}$. This low-rank approximation has been shown to yield more robust performance in scenarios where accurate estimation of $\mathbf{R}_\mathbf{ss}$ is challenging.

Since the \mbox{\gls{tidanse}} algorithm is also based on the signal model from~(\ref{eq:signalmodel}), $\tilde{\mathbf{R}}_{\mathbf{s}_k\mathbf{s}_k}^i$ is also rank-$S$. This can be seen from the definitions below~(\ref{eq:mwf_danse}), rewriting:

\begin{gather}
    \tilde{\mathbf{R}}_{\mathbf{s}_k\mathbf{s}_k}^i
    = \mathbb{E}\left\{
        \tilde{\mathbf{s}}_k^i\tilde{\mathbf{s}}_k^{i\Her}
    \right\}
    = \mathbb{E}\left\{
        \mathbf{C}_k^{i\Her}\mathbf{s}\mathbf{s}^\Her\mathbf{C}_k^{i}
    \right\}
    = \mathbf{C}_k^{i\Her} \mathbf{R}_{\mathbf{ss}} \mathbf{C}_k^{i},\\
    \text{where:}\:\:\:
    \mathbf{C}_k^{i\Her} = \begin{bmatrix}
        \mathbf{0} \:\:\:\: \cdots \:\:\:\:\: \mathbf{0} \:\:\:\: \mathbf{I}_{M_k} \:\:\: \mathbf{0} \:\:\:\: \cdots \:\:\:\: \mathbf{0}\\
        \mathbf{P}_1^{i\Her} \: \cdots \: \mathbf{P}_{k-1}^{i\Her} \:\: \mathbf{0} \:\:\: \mathbf{P}_{k+1}^{i\Her} \: \cdots \: \mathbf{P}_{K}^{i\Her}
    \end{bmatrix}.
\end{gather}

Since $\mathbf{R}_{\mathbf{ss}}$ is rank-$S$, so is $\tilde{\mathbf{R}}_{\mathbf{s}_k\mathbf{s}_k}^i$. This low-rank property can be guaranteed by first making use of a \gls{gevd} on the pencil $\{
    \tilde{\mathbf{R}}_{\mathbf{y}_k\mathbf{y}_k}^i,
    \tilde{\mathbf{R}}_{\mathbf{n}_k\mathbf{n}_k}^i
\}$ to rewrite the \glspl{scm} as:

\begin{equation}\label{eq:GEVDoutcome}
    \tilde{\mathbf{R}}_{\mathbf{y}_k\mathbf{y}_k}^i
    = \tilde{\mathbf{Q}}_k^i \mathbf{\Sigma}_k^i \tilde{\mathbf{Q}}_k^{i\Her}
    \:\:\:\text{and}\:\:\:
    \tilde{\mathbf{R}}_{\mathbf{n}_k\mathbf{n}_k}^i
    = \tilde{\mathbf{Q}}_k^i \tilde{\mathbf{Q}}_k^{i\Her},
\end{equation}

\noindent
where $\mathbf{\Sigma}_k^i = \mathrm{diag}\{
    \tilde{\sigma}_{k1}^i,...,\tilde{\sigma}_{kD_k}^i
\}$ contains the generalized eigenvalues ordered from largest to smallest and $\tilde{\mathbf{Q}}_k^i$ contains the corresponding generalized eigenvectors.
The rank can be constrained to $R$ (ideally equal to $S$, when $S$ is known a priori) by setting the $D_k - R$ smallest eigenvalues in $\tilde{\mathbf{R}}_{\mathbf{y}_k\mathbf{y}_k}^i$ to zero, obtaining an estimate of $\tilde{\mathbf{R}}_{\mathbf{s}_k\mathbf{s}_k}$ as:

\begin{equation}\label{eq:RssGEVDest}
    \hat{\tilde{\mathbf{R}}}_{\mathbf{s}_k\mathbf{s}_k}
    = \hat{\tilde{\mathbf{R}}}_{\mathbf{y}_k\mathbf{y}_k}^i - \tilde{\mathbf{R}}_{\mathbf{n}_k\mathbf{n}_k}^i
    =
    \tilde{\mathbf{Q}}_k^i \mathbf{\Delta}_k^i \tilde{\mathbf{Q}}_k^{i\Her},
\end{equation}

\noindent
where $\mathbf{\Delta}_k^i = \mathrm{diag}\{
    \tilde{\sigma}_{k1}^i-1, \dots, \tilde{\sigma}_{kR}^i-1, 0, \dots, 0
\}$.
This results in the \mbox{\gls{tigevddanse}} algorithm, which solution to~(\ref{eq:mmseTIDANSE}) is obtained by substituting (\ref{eq:RssGEVDest}) into (\ref{eq:mwf_danse}):

\begin{equation}\label{eq:filterUpdateTIGEVDDANSE}
    \tilde{\mathbf{W}}_k^{i+1}
    =
    (\tilde{\mathbf{Q}}_k^i)^{-\Her}\tilde{\mathbf{\Lambda}}_k^i\tilde{\mathbf{Q}}_k^{i\Her}\tilde{\mathbf{E}}_{k},
\end{equation}

\noindent
where $\tilde{\mathbf{\Lambda}}_k^i = \mathrm{diag}\{
    1-1/\tilde{\sigma}_{k1}^i \dots 1-1/\tilde{\sigma}_{kR}^i, 0\dots 0
\}$. The rest of the algorithm remains unchanged with respect to~Section~\ref{sec:TIDANSE}. In addition to enabling enhanced and more robust signal estimation performance, it is observed that this rank-$R$ approximation preserves convergence of \gls*{tigevddanse} even in cases where $J$ or $R$ underestimates the number of latent desired sources $S$. This is not the case of \gls*{tidanse} which relies on the assumption $S\leq J$. Although a convergence proof is not provided in this paper, the observed property is illustrated via the experimental results showcased in~Section~\ref{sec:numericalExperiments}.

%%%%%
\section{Improved Robustness}\label{sec:improvedRobustness}
%%%
\subsection{Non-Strict Convergence of \mbox{\gls{ti_or_gevddanse}}}

It is observed that the \mbox{\gls{ti_or_gevddanse}} algorithm exhibits a non-strict convergence due to the formulation of its fusion matrices.
As derived in Section~\ref{sec:TIDANSE}, $\tilde{\mathbf{y}}_k^i = \mathbf{C}_k^{i\Her}\mathbf{y}$, implying that the desired signal estimate can be written as $\hat{\mathbf{d}}_k^{i+1} = \tilde{\mathbf{W}}_k^{i+1,\Her}\tilde{\mathbf{y}}_k^i
= \tilde{\mathbf{W}}_k^{i+1,\Her}\mathbf{C}_k^{i\Her}\mathbf{y}$.
Therefore, the $M \times J$ matrix $\mathbf{W}_k^{i+1} = \mathbf{C}_k^{i}\tilde{\mathbf{W}}_k^{i+1}$ is the network-wide version of the \mbox{\gls{ti_or_gevddanse}} filter matrix, with structure:

\begin{equation}\label{eq:netWideTIDANSEmat}
    \mathbf{W}_k^{i+1}
    =\begin{bmatrix}
        \left[
            (\mathbf{P}_1^{i}\mathbf{G}_k^{i+1})^\T\:|\:
            \cdots\:|\:
            (\mathbf{P}_{k-1}^{i}\mathbf{G}_k^{i+1})^\T
        \right]^\T\\
        \mathbf{W}_{kk}^{i+1}\\
        \left[
            (\mathbf{P}_{k+1}^{i}\mathbf{G}_k^{i+1})^\T\:|\:
            \cdots\:|\:
            (\mathbf{P}_{K}^{i}\mathbf{G}_k^{i+1})^\T
        \right]^\T
    \end{bmatrix}.
\end{equation}

Inspecting the terms $\mathbf{P}_q^{i}\mathbf{G}_k^{i+1} = \mathbf{W}_{qq}^{i}(\mathbf{G}_q^{i})^{-1}\mathbf{G}_k^{i+1}$ in (\ref{eq:netWideTIDANSEmat}) reveals that convergence of $\mathbf{W}_k^i$ (and thus $\tilde{\mathbf{W}}_k^i$) does not require the filter coefficients in the $\{\mathbf{G}_k^i\}_{k\in\mathcal{K}}$ matrices to strictly converge. In fact, only the strict convergence of $(\mathbf{G}_q^i)^{-1}\mathbf{G}_k^{i+1}\:\forall\:(k,q)\in\mathcal{K}\times\mathcal{K}\backslash\{k\}$ is necessary. In this state, the \mbox{\gls{ti_or_gevddanse}} algorithm allows the elements of the $\{\mathbf{G}_k^i\}_{k\in\mathcal{K}}$ matrices to grow infinitely large or small as $i$ increases. Such behavior is indeed observable in practice and can lead to numerical overflow or significant precision errors.

%%%%
\subsection{Normalization strategy}

In this section, we address the non-strict convergence of \mbox{\gls{ti_or_gevddanse}}.
Let us define a normalization factor $\gamma^i\in\mathbb{C}$ and suppose that every node simultaneously start normalizing their $\mathbf{G}_k^i$ matrix at iteration $i$. The normalized fused signal $\bar{\mathbf{z}}_k^i$ can be related to its non-normalized counterpart via:

\begin{equation}\label{eq:z_norm}
    \bar{\mathbf{z}}_k^{i} =
    (
        \mathbf{W}_{kk}^{i}(
            \mathbf{G}_k^{i}
        )^{-1}
        \gamma^{i}
    )^\Her
    \mathbf{y}_k
    =
    \gamma^{i,\ast}
    \mathbf{P}_k^{i\Her}
    \mathbf{y}_k
    =
    \gamma^{i,\ast}
    \mathbf{z}_k^i.
\end{equation}

\noindent
where $\cdot^\ast$ denotes the complex conjugate.
Following the same notation logic, $\bar{\boldsymbol{\eta}}^i = \sum_{k\in\mathcal{K}} \bar{\mathbf{z}}_k^{i} = \gamma^{i,\ast}\boldsymbol{\eta}^i$, meaning that the normalized version of $\tilde{\mathbf{y}}_k^i$ can be expressed as $\bar{\tilde{\mathbf{y}}}_k^i = \mathbf{N}_k^i\tilde{\mathbf{y}}_k^i$ where $\mathbf{N}_k^i = \mathrm{blkdiag}\{\mathbf{I}_{M_k}, \gamma^i\mathbf{I}_{J}\}$.
It follows that: 

\begin{equation}
    \bar{\tilde{\mathbf{R}}}_{\mathbf{y}_k\mathbf{y}_k}^i
    = \mathbb{E}\left\{\bar{\tilde{\mathbf{y}}}_k^i\bar{\tilde{\mathbf{y}}}_k^{i\Her}\right\}
    = \mathbf{N}_k^i\tilde{\mathbf{R}}_{\mathbf{y}_k\mathbf{y}_k}^i\mathbf{N}_k^{i\Her},
\end{equation}

\noindent
and likewise for $\bar{\tilde{\mathbf{R}}}_{\mathbf{n}_k\mathbf{n}_k}^i$. Substituting in (\ref{eq:GEVDoutcome}) gives:

\begin{equation}
    \bar{\tilde{\mathbf{R}}}_{\mathbf{y}_k\mathbf{y}_k}^i
    = \bar{\tilde{\mathbf{Q}}}_k^i \mathbf{\Sigma}_k^i \bar{\tilde{\mathbf{Q}}}_k^{i\Her}
    \:\:\:\text{and}\:\:\:
    \bar{\tilde{\mathbf{R}}}_{\mathbf{n}_k\mathbf{n}_k}^i
    = \bar{\tilde{\mathbf{Q}}}_k^i \bar{\tilde{\mathbf{Q}}}_k^{i\Her},
\end{equation}

\noindent
where $\bar{\tilde{\mathbf{Q}}}_k^i = \mathbf{N}_k^i\tilde{\mathbf{Q}}_k^i$. Consequently, (\ref{eq:filterUpdateTIGEVDDANSE}) gives:

\begin{equation}
    \bar{\tilde{\mathbf{W}}}_k^{i+1}
    =
    (\bar{\tilde{\mathbf{Q}}}_k^i)^{-\Her}\tilde{\mathbf{\Lambda}}_k^i\bar{\tilde{\mathbf{Q}}}_k^{i\Her}\tilde{\mathbf{E}}_{k}
    =
    (\mathbf{N}_k^i)^{-\Her}
    \tilde{\mathbf{W}}_k^{i+1},
\end{equation}

\noindent
where use is made of the fact that the desired signal only includes contributions from local sensors, i.e., $\mathbf{N}_k^{i\Her}
\tilde{\mathbf{E}}_{k} = \tilde{\mathbf{E}}_{k}$.
It can be shown that the normalization does not alter the network-wide filters of (\ref{eq:netWideTIDANSEmat}). This can be seen from:

\begin{equation}\label{eq:d_withNorm}
    \bar{\hat{\mathbf{d}}}_k^{i+1} = \bar{\tilde{\mathbf{W}}}_k^{i+1,\Her}\bar{\tilde{\mathbf{y}}}_k^i
    =
    \tilde{\mathbf{W}}_k^{i+1,\Her}
    (\mathbf{N}_k^i)^{-1}
    \mathbf{N}_k^i
    \tilde{\mathbf{y}}_k^i
    = \hat{\mathbf{d}}_k^{i+1}.
\end{equation}

This normalization procedure can be incorporated in \mbox{\gls{tigevddanse}}, resulting in the following algorithm:

\begin{algorithmic}[1]
    \STATE Initialize $u=0$, $\gamma^{1} = 1$, $r\in\mathcal{K}$.
    \FOR{$i=1,2,3,\dots$}
        \STATE Form tree topology rooted at node $u$.
        \STATE At all $k\in\mathcal{K}$, compute $\bar{\mathbf{z}}_k^{i} = \gamma^{i,\ast} \bar{\mathbf{P}}_k^{i\Her} \mathbf{y}_k$.
        \STATE At node $u$, compute the in-network sum $\boldsymbol{\eta}^{i}$ and flood it back through the \gls{wasn}.
        \FOR{$k\in\mathcal{K}$}
            \STATE Build observation vector $\bar{\tilde{\mathbf{y}}}_k^i$ as in (\ref{eq:yTildeTI}).
            \STATE Compute \glspl{scm} $\bar{\tilde{\mathbf{R}}}_{\mathbf{y}_k\mathbf{y}_k}^{i}$ and $\bar{\tilde{\mathbf{R}}}_{\mathbf{n}_k\mathbf{n}_k}^{i}$.
            \IF{$k=u$}
                \STATE Compute $\bar{\tilde{\mathbf{W}}}_k^{i+1}=[\bar{\mathbf{W}}_{kk}^{i+1,\T}\:|\:\bar{\mathbf{G}}_{k}^{i+1,\T}]^\T$ via (\ref{eq:filterUpdateTIGEVDDANSE}).
            \ELSIF{$k\neq u$}
                \STATE Compute $\bar{\tilde{\mathbf{W}}}_k^{i+1} = \left(\mathbf{N}_k^i\right)^{-\Her} \bar{\tilde{\mathbf{W}}}_k^{i}$.
            \ENDIF
            \STATE Update fusion matrix as $\bar{\mathbf{P}}_k^{i+1} = \bar{\mathbf{W}}_{kk}^{i+1}(\bar{\mathbf{G}}_{k}^{i+1})^{-1}$.
        \ENDFOR
        \STATE At reference node $r$, compute $\gamma^{i+1} = \|\bar{\mathbf{G}}_r^{i+1}\|_F$ and flood it through the \gls*{wasn}.
        \STATE At all $k\in\mathcal{K}$, compute $\hat{\mathbf{d}}_k^{i+1} =
        \bar{\tilde{\mathbf{W}}}_k^{i+1,\Her}\bar{\tilde{\mathbf{y}}}_k^{i+1}$.
        \STATE $u \gets (u + 1) \mod K$.
    \ENDFOR
\end{algorithmic}

\vspace*{.5em}
\noindent
\textbf{Remark 1: }
The choice for $\gamma^i$ is to use $\|\bar{\mathbf{G}}_r^{i+1}\|_F$ for a fixed reference node $r$ (step 15). It should be noted that the algorithm can in principle select any node $r$ as reference, as long as the normalization is equal for all nodes.

\vspace*{.5em}
\noindent
\textbf{Remark 2: }
Step 12 ensures consistency across iterations for non-updating nodes $k\neq u$. Indeed, the $\gamma^{i,\ast}$ factor from step 4 is carried along in the \glspl{scm} and thus impacts the filter of the updating node. Conversely, a non-updating node will keep its filters from the previous iteration, meaning that the normalization effect must be accounted for via $(\mathbf{N}_k^i)^{-\Her}$.

\vspace*{.5em}
\noindent
\textbf{Remark 3: } The communication bandwidth increase generated by the broadcasting of $\gamma^{i+1}$ (step 15) is negligible in comparison to the unnormalized \mbox{\gls{tigevddanse}} algorithm, as it represents the exchange of a single scalar at most at every iteration. Depending on the severity of the non-strictly convergent behavior of the $\{\bar{\mathbf{G}}_k^i\}_{k\in\mathcal{K}}$ coefficients, it may even be sufficient to update (and thus broadcast) $\gamma$ less frequently.

\vspace*{.5em}
When estimating \glspl{scm} in an online fashion based on~(\ref{eq:scmsOnline}), the normalization must be accounted for by setting: 

\begin{equation}\label{eq:scmsOnlineWithNorm}
    \begin{split}
        \bar{\tilde{\mathbf{R}}}_{\mathbf{y}_k\mathbf{y}_k}^i
        &=
        \beta
        \mathbf{N}_k^i \bar{\tilde{\mathbf{R}}}_{\mathbf{y}_k\mathbf{y}_k}^{i-1} \mathbf{N}_k^{i\Her}
        + (1 - \beta)
        \bar{\tilde{\mathbf{y}}}_k^i\bar{\tilde{\mathbf{y}}}_k^{i\Her},\\
        \bar{\tilde{\mathbf{R}}}_{\mathbf{n}_k\mathbf{n}_k}^i
        &=
        \beta
        \mathbf{N}_k^i \bar{\tilde{\mathbf{R}}}_{\mathbf{n}_k\mathbf{n}_k}^{i-1} \mathbf{N}_k^{i\Her}
        + (1 - \beta)
        \bar{\tilde{\mathbf{y}}}_k^i\bar{\tilde{\mathbf{y}}}_k^{i\Her}.
    \end{split}
\end{equation}

\noindent
where the index $i$ here represents both the iteration index and the time index, for simplicity. 

%%%%%
\section{Numerical Experiments}\label{sec:numericalExperiments}

The performance of \mbox{\gls{tigevddanse}} with and without normalization is assessed via simulations in an acoustic environment composed of an ad-hoc non-fully connected \gls{wasn} and 6 localized sources, $S{=}3$ of which are considered as targets and the $S_n{=}3$ others as noise.
The number of nodes is fixed but the specific \gls{wasn} topology does not have to be, as it is simply assumed that $\boldsymbol{\eta}^i$ is available at all nodes at any $i$.
Although the \gls{gevd}-rank $R$ can be chosen independently from the number of exchanged channels $J$, we here set $R{=}J$ for simplicity (other cases are discussed in, e.g.,~\cite{hassaniGEVD}).
The uncorrelated (thermal) noise at node $k$ is set to have a power equal to 10\% of the power of the combined target source signals as observed by the first sensor of the node. Signal samples and steering matrices entries are drawn from the uniform distribution over \mbox{[-0.5, 0.5]}.

%%%
\subsection{Batch-mode simulations}

\begin{figure}[h]
    \centering
    \begin{scriptsize}
        \input{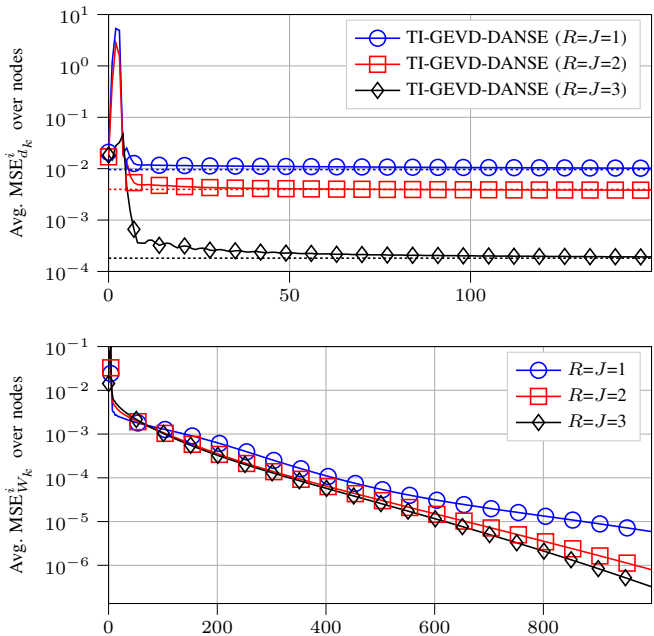}
    \end{scriptsize}
    \caption{Batch processing results. The $x$-axis represents the iteration index $i$. Top: $\mathrm{MSE}_{d_k}^i$ averaged over nodes, where the horizontal dotted lines represent the corresponding centralized values. Bottom: $\mathrm{MSE}_{W_k}^i$ averaged over nodes.}
    \label{fig:batch_res}
\end{figure}

Batch-mode simulations without normalization ($\gamma^i = 1\:\forall\:i$) are first performed with $K{=}5$ and $M_k{=}4$ for all $k$ to demonstrate the convergence of \mbox{\gls{tigevddanse}} towards the centralized \mbox{\gls{gevdmwf}}. The number of observations is set to $N{=}15000$ samples. The \glspl{scm} are estimated as sample means. The results are averaged over 3 runs with different random steering matrices.
In order to isolate \gls{scm} estimation errors, $\mathbf{R}_\mathbf{nn}$ and $\bar{\tilde{\mathbf{R}}}^i_{\mathbf{n}_k\mathbf{n}_k}$ are estimated using oracle knowledge of the noise-only signals.
The performance of the \mbox{\gls{tigevddanse}} algorithm is shown for different values of $R{=}J$ in Fig.~\ref{fig:batch_res}, where two \gls{mse} metrics are used. The first, $\mathrm{MSE}_{W_k}^i$, is defined between the network-wide expansion of the \mbox{\gls{tigevddanse}} solution of~(\ref{eq:netWideTIDANSEmat}) and the centralized \mbox{\gls{gevdmwf}} solution of~(\ref{eq:mmseCentr}). The second, $\mathrm{MSE}_{d_k}^i$, is defined between the true desired signal and its estimate $\hat{\mathbf{d}}_k^i$:

\begin{align}
    \mathrm{MSE}_{W_k}^i &= \frac{1}{MJ}\|
        \bar{\mathbf{W}}_k^i - \mathbf{W}_k
    \|_F^2,\\
    \mathrm{MSE}_{d_k}^i &= \frac{1}{JN}\sum_{n=0}^{N-1}
    \|
        \hat{\mathbf{d}}_k^i[n] - \mathbf{d}_k[n]
    \|_F^2.
\end{align}

The convergence of the \mbox{\gls{tigevddanse}} algorithm towards the centralized \mbox{\gls{gevdmwf}} is clearly visible from the average $\mathrm{MSE}_{W_k}^i$, which would decrease to machine precision for $i{\to}\infty$. Larger values of $R{=}J$ show a faster convergence since more distinct signals are available at each nodes. It is important to notice that convergence is preserved even in cases where $R{=}J{<}S$, which is not the case for the original \gls*{tidanse}~\cite{szurleyTopology}. The average $\mathrm{MSE}_{d_k}^i$ shows that \mbox{\gls{tigevddanse}} also matches the target signal estimation performance of the \mbox{\gls{gevdmwf}} with corresponding rank, and that a rank $R$ closer (or equal) to $S$ yields better performance.

%%%
\subsection{Online-mode simulations}

The behavior of \mbox{\gls{tigevddanse}} is assessed in an online processing scenario with and without normalization, with $K{=}10$ and $M_k{=}15$ for all $k$. The \gls{scm} estimation strategy defined in~(\ref{eq:scmsOnlineWithNorm}) is used with $\beta{=}0.7$ and all \gls{scm} entries randomly initialized.
At every iteration, a new frame of $B=500$ samples is drawn from the uniform distribution over \mbox{[-0.5, 0.5]} for each sound source.
A dynamic scenario is created where the entries of steering matrices $\mathbf{A}$ and $\mathbf{B}$ defined in~(\ref{eq:signalmodel}) have, at each time frame, a 0.05 probability to be changed by drawing entries from the same uniform distribution over \mbox{[-0.5, 0.5]}, thus changing the relative positioning of sensors and sources. After an acoustic scenario change, the probability is set to 0 for 30 frames before coming back to 0.05.

\begin{figure}[h]
    \centering
    \begin{scriptsize}
        \input{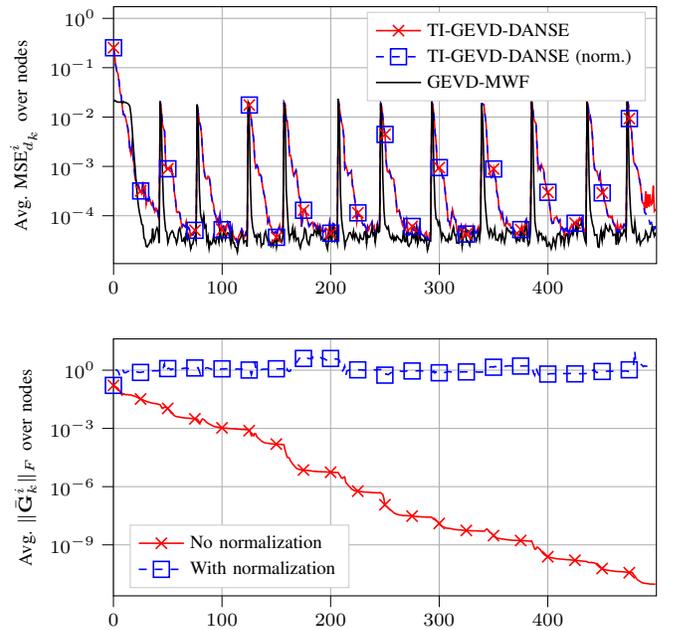}
    \end{scriptsize}
    \caption{Online processing results. The $x$-axis represents the index $i$. Top: $\mathrm{MSE}_{d_k}^i$ averaged over nodes. Bottom: $\|\bar{\mathbf{G}}_k^i\|_F$ averaged over nodes.}
    \label{fig:online_res}
\end{figure}

The quantity $\mathrm{MSE}_{d_k}^i$ averaged over all nodes is shown in~Fig.~\ref{fig:online_res} for \mbox{\gls{tigevddanse}} with and without normalization and for the centralized \mbox{\gls{gevdmwf}}, all with $J{=}R{=}S{=}3$.
To highlight the effect of $\gamma^i$, the average over all nodes of $\|\bar{\mathbf{G}}_k^{i}\|_F$ is also shown for \mbox{\gls{tigevddanse}} with and without normalization. The reference node index $r$ is kept equal to 1 through the entire simulation.

The results show that, as long as \mbox{\gls{tigevddanse}} is stable without normalization, the normalization has no impact on the performance as per~(\ref{eq:d_withNorm}), even in a dynamic scenario. It can be noticed that a change in the steering matrices triggers a new adaptation phase where the algorithm must \mbox{re-estimate} the \glspl{scm} to be able to \mbox{re-converge} towards the centralized \mbox{\gls{gevdmwf}} (visible as peaks in the average $\mathrm{MSE}_{d_k}^i$).

The importance of normalization is clearly visible on the lower plot. Without normalization, the non-strict convergence of \mbox{\gls{tigevddanse}} leads to a decreasing $\|\bar{\mathbf{G}}_k^i\|_F = \|\mathbf{G}_k^i\|_F$. Steeper decreases occur when the acoustic scenario is changed, as the \glspl{scm} must be re-estimated. This behavior results in numerical instability over time as can be seen from $i=488$ in the upper plot. With normalization, however, the average norm of $\bar{\mathbf{G}}_k^i$ remains close to 1 even if the acoustic scenario changes, which allows \mbox{\gls{tigevddanse}} to indefinitely perform without numerical overflow or critical precision errors.

\section{Conclusion}\label{sec:conclusion}

In this paper, we have introduced a \gls{gevd}-based \mbox{\gls{tidanse}} algorithm (\mbox{\gls{tigevddanse}}) which provides an enhanced and robust performance in scenarios where the accurate estimation of the desired-signal \gls{scm} is challenging. An adaptive normalization procedure has been included, ensuring the stability of the estimated filter coefficients through time, even in online processing with time-varying acoustic scenarios.

\end{document}